\documentclass{article}

\usepackage{arxiv}
\usepackage[utf8]{inputenc} 
\usepackage[T1]{fontenc}    
\usepackage{hyperref}       
\usepackage{url}            
\usepackage{booktabs}       
\usepackage{amsfonts}       
\usepackage{nicefrac}       
\usepackage{microtype}
\usepackage{lipsum}		
\usepackage{graphicx}
\usepackage{doi}
\usepackage{listings}
\usepackage[font=normalsize,labelfont={bf}]{caption}
\usepackage{upquote}

\hypersetup{
    colorlinks=true,
    linkcolor=blue,
    urlcolor=blue,
}

\graphicspath{ {./images/} }

\title{An Analysis into the Performance and Memory Usage of MATLAB Strings}

\date{June 12\textsuperscript{th}, 2020}

\author{ {\hspace{1mm}Travis Near} \\
	Department of Computer Science \\
	Missouri University of Science and Technology \\
	Rolla, MO 65409 \\
	\texttt{tentxb at mst.edu} \\
}

\renewcommand{\headeright}{}
\renewcommand{\undertitle}{}

\begin{document}
\maketitle

\begin{abstract}
MATLAB is a mathematical computing environment used by many engineers, mathematicians, and students to process and understand their data. Important to all data science is the managing of textual data. MATLAB supports two textual data containers: (1) \emph{cell arrays} of characters and (2) \emph{string arrays}. This research showcases the strengths of string arrays over cell arrays by quantifying their performance, memory contiguity, syntax readability, interface fluidity, and autocomplete capabilities. These results demonstrate that string arrays often run 2x to 40x faster than cell arrays for common string benchmarks, are optimized for data locality by reducing metadata overhead, and offer a more expressive syntax due to their automatic data type conversions and vectorized methods.
\end{abstract}


\section{Introduction}
MATLAB has supported text since its inception as a simple array of numeric ASCII values \cite{Moler}. Textual data in programming languages is a simple concept but expands into many implementation challenges regarding its performance and memory layout. This analysis begins by comparing common usages of two different containers of text in MATLAB: \textbf{\emph{cell array}} and \textbf{\emph{string array}}. Next, it measures these containers' performance by running integral string benchmarks: string building, formatting, data type conversions, and concatenation. Beyond performance, this paper also quantifies cell array and string array ease of use and interface conciseness. Then it details the difficult-to-measure metadata allocation of text to shed light on these containers' internal memory structure. Lastly, this research closes with a glimpse into string compilation and how to beat MATLAB string speed for duration-critical operations.

\section{Different text types in MATLAB}
\subsection{Char vector}
Text is composed of characters and each character in MATLAB is two bytes in size. A \href{https://www.mathworks.com/help/matlab/ref/char.html}{\emph{char vector}} is a simple horizontal sequence of characters. \emph{\textquotesingle hello\textquotesingle}, for example, is a char vector. MATLAB denotes \emph{char} with single quotes.

\subsection{Char matrix}
A \emph{char matrix} is an \emph{MxN} matrix of characters. Unlike with double-precision matrices, sets of characters are rarely perfectly rectangular: almost all collections of text are of differing lengths. The example below demonstrates how to construct a char matrix. Note how the shorter string must be padded with spaces to create a rectangular matrix:
\begin{verbatim}
    >> text = ['hello   '; 'everyone']
text =

  2×8 char array

    'hello   '
    'everyone'
\end{verbatim}

In addition to requiring padding, char matrix is unnatural due to MATLAB's column-major layout \cite{ColumnMajor}. The \emph{\textquotesingle hello everyone\textquotesingle} text above is stored in columns in memory as \emph{\textquotesingle heevlelroy o n e\textquotesingle} (indexes 1, 3, 5, 7 correspond to \emph{\textquotesingle hello\textquotesingle} and indexes 2, 4, 6, ..., and 16 correspond to \emph{\textquotesingle everyone\textquotesingle}). Even though strings in English read left-to-right, data in MATLAB is always stored in vertical columns. Iterating by rows in MATLAB is difficult and yields poor performance due to jumping forward and back across the contiguous data buffer. Given these limitations of char matrix, developers typically use a higher-level container to hold their textual data such as a \emph{cell array of character vectors}.

\subsection{Cell array of character vectors (\emph{cellstr})}
The classic container to house collections of text in MATLAB is \emph{cell array}. Although textual data is uniform in type, it is frequently stored in a non-uniform cell array (see \S \ref{DataUniformity} for data uniformity). A cell array of character vectors is also known as a \emph{cellstr} (short for \emph{cell string}). One cell array advantage over char matrix is that it can hold text of differing number of characters. \emph{cellstr} is created with braces (\{\}) and single quotes (\textquotesingle). Here is the same \emph{\textquotesingle hello everyone\textquotesingle} text stored in a cell array without needing additional padding:
\begin{verbatim}
>> cell_array = {'hello'; 'everyone'}
cell_array =

  2×1 cell array

    {'hello'   }
    {'everyone'}
\end{verbatim}

\subsection{String}
In MATLAB, \href{https://www.mathworks.com/help/matlab/ref/string.html}{\emph{string}} is a much newer data type compared to \emph{char} and \emph{cell}. MATLAB uses double quotes to denote strings (e.g., \emph{"world"}). In terms of MATLAB syntax, \emph{char} and \emph{string} are distinguished only by their quote character: single quote for \emph{char} and double quote for \emph{string}. Internally, however, string has a distinct data structure which is conducive to creating powerful \emph{string arrays}.

\subsection{String array}
MATLAB introduced string array in R2016b to help supersede \emph{cellstr} \cite{Cellstr}. Like cell array, string arrays allow holding strings of differing lengths. String arrays are created using brackets ([]) and double quotes ("):
\begin{verbatim}
    >> string_array = ["hello"; "everyone"]
string_array = 

  2×1 string array

    "hello"
    "everyone"
\end{verbatim}

Cosmetically, string array and \emph{cellstr} are similar, but analytically the differences are huge. Section \ref{StringComparison} analyzes their differences in execution speed, memory usage, data uniformity, and clarity.

\section{Cell array of character vectors vs. string array}
\label{StringComparison}

The difference between \emph{cellstr} and \emph{string array} is best introduced through an example. Consider this exercise which will first be solved using \emph{cellstr} then later using string array: \\

\emph{Using MATLAB, generate a sequence of the strings "TestResult1", "TestResult2", ..., "TestResult1000".} \\

\subsection{Cell array solution}
Using a cell array of character vectors, here is a MATLAB one-liner to generate the \emph{"TestResult\#"} strings:

\begin{verbatim}
    >> c = arrayfun(@(idx) ['TestResult', num2str(idx)], 1:1000, 'UniformOutput', false);
\end{verbatim}

This complex line has much to dissect:

\begin{itemize}
    \item \emph{arrayfun} to expand to 1000 elements
    \item \emph{num2str} to convert each numeric index to its ASCII \emph{char} counterpart
    \item Brackets ([]) to concatenate \emph{\textquotesingle TestResult\textquotesingle} with the converted index
    \item \emph{UniformOutput} set to false to create a \emph{cell array} (char vectors of differing number of elements are \emph{not} uniform, contrasted with a char \emph{matrix} which is uniform, see \S \ref{DataUniformity})
\end{itemize}

Much of this syntax is unique to MATLAB. Experienced MATLAB users understand these details, but new users who have a Python or non-development background might find them intimidating: \emph{What is arrayfun? Why does num2str return char instead of string? What is UniformOutput? ...and why don't I want it?}

\subsection{String array solution}
The MATLAB expression below also creates the \emph{"TestResult\#"} 1000-element sequence in one line of M-code, but this time using double quoted \emph{string arrays}:

\begin{verbatim}
    >> s = "TestResult" + (1:1000);
\end{verbatim}

Unlike with \emph{cellstr}, this line reads like a natural sentence: no function calls, no concerns over data uniformity, and a familiar string concatenation `+' operator. Aesthetics aside, these two lines can be compared quantitatively as shown in Table \ref{table:StringBuilding}. \\

\begin{table}[ht]
\begin{tabular}{|l|l|l|l|}
\texttt{Metric}               & \texttt{c = arrayfun(@(idx)}          & \texttt{s = "TestResult" +}          & \texttt{String advantage} \\ \hline
\texttt{Characters of M-code} & \texttt{78 chars}                     & \texttt{24 chars}                    & \texttt{3.25x shorter}    \\
\texttt{Duration (sec)}       & \texttt{0.01640}                      & \texttt{0.0003634}                   & \texttt{45x faster}       \\
\texttt{Bytes}                & \texttt{129,786}                      & \texttt{70,096}                      & \texttt{1.85x smaller}   
\end{tabular}
\caption{Performance comparison of string building for cell and string array}
\label{table:StringBuilding}
\end{table}

This table begins to unlock some of the key differences between cell array and string array:

\underline{Character count}

78 characters of MATLAB code for \emph{cellstr} compared to 24 characters for \emph{string}. Removing the boilerplate code needed for \emph{cellstr} distills the \emph{string} expression into its constituent text and numeric elements. \\

\underline{Performance}

0.01640 seconds for \emph{cellstr} vs. 0.0003634 seconds for \emph{string}, giving string a 45x speedup. Function calls and loops (\emph{arrayfun} is a loop masquerading as a function call) have significant cost compared to built-in vectorization. \\

\underline{Memory usage}

129,786 bytes vs. 70,096 bytes. Understanding why string array requires less memory requires additional explanation: \\

\subsection{Memory layout comparison of cell array and string array}

The \emph{cellstr} in Table \ref{table:StringBuilding} is 129,786 bytes while the string array is only 70,096 bytes even though their textual content is identical: these data types differ in how they require \emph{metadata}. Knowing the byte size of the raw text allows calculating the metadata size that each data type needs (characters in MATLAB are \textbf{2 bytes} each):

\begin{verbatim}
    Vector to analyze: ["TestResult1", "TestResult2", ..., "TestResult1000"].
    
    'TestResult' text = 10 chars * 2 bytes/char * 1000 copies = 20,000 bytes.
\end{verbatim}

Then account for the numeric suffixes of each "TestResult\#":

\begin{verbatim}
    1-9     = 1 digit  * 2 bytes/char *   9 elements =   18 bytes
    10-99   = 2 digits * 2 bytes/char *  90 elements =  360 bytes
    100-999 = 3 digits * 2 bytes/char * 900 elements = 5400 bytes
    1000    = 4 digits * 2 bytes/char *   1 element  =    8 bytes
\end{verbatim}

Summing it together, the total bytes of raw data is 
    \(20000 + 18 + 360 + 5400 + 8 = \textbf{25,786}\) bytes.\newline
    
MATLAB can also derive the number of data bytes (25,786) by joining all the discrete strings into one long character vector:
\begin{verbatim}
    >> s = "TestResult" + (1:1000);
    >> raw = s.join('').char;
    >> whos raw

      Name      Size       Bytes
      raw       1x12893    25786
\end{verbatim}

Knowing the total bytes and raw data bytes yields the amount of metadata which \emph{cellstr} and \emph{string array} need as summarized in Table \ref{table:MetadataBytes}.

\begin{table}[ht]
\begin{tabular}{|l|l|l|l|}
\texttt{Metric}       & \texttt{Total bytes} & \texttt{Raw data bytes} & \texttt{Metadata (total - raw)}        \\ \hline
\texttt{cellstr}      & \texttt{129786}      & \texttt{25786}          & \texttt{104000 (2.34x more bloated)  } \\
\texttt{string array} & \texttt{70096}       & \texttt{25786}          & \texttt{44310}                        
\end{tabular}
\caption{Comparison of data and metadata sizes of cell array and string array}
\label{table:MetadataBytes}
\end{table}

Based on Table \ref{table:MetadataBytes}, \emph{cellstr} requires 2.34x (\(104000 / 44310 = 2.34\)) as much metadata to store the same "TestResult\#" array. This metadata difference is primarily due to \emph{cellstr} needing 1000 \emph{mxArrays} and \emph{string array} needing only 1 \emph{mxArray}.

\subsection{\emph{mxArray}}
\label{mxArray}
\emph{mxArray} is the fundamental C++ data type for all MATLAB variables. \emph{mxArray} is implemented as a \emph{tagged union} \cite{MATLABData}. Tagged unions strike a balance: they minimize byte allocation while maximizing support for distinct data types. This allows the \emph{mxArray} to be flexible: it supports numeric, text, sparse, struct, cell, etc. The trade-off is that the \emph{mxArray} is comparatively large due to MATLAB's language complexity: 104 bytes for each \emph{mxArray}. This number of bytes (104) can be derived by creating a cell array husk:
\begin{verbatim}
    >> c = {[]};
    >> whos c
      Name      Size      Bytes
      c         1x1         104
\end{verbatim}

Cell arrays by definition are non-uniform and thus need one \emph{mxArray} per element \cite{CellArray}. Therefore, a 1000-length cell array needs 1000 \emph{mxArrays}, totaling 1000 * 104 bytes = 104,000 bytes of metadata. String arrays, by contrast, are uniform and thus only require one \emph{mxArray} regardless of the string array size. Its smaller overhead (44,310 bytes, from Table \ref{table:MetadataBytes}) stems from its internal small vector container in C++ which annotates the size of the textual content.

\subsection{Data uniformity}
\label{DataUniformity}
Data in MATLAB is considered \emph{uniform} when a function called on each element returns a \emph{scalar} which can be concatenated into an array \cite{Uniform}. For example, \([1]\), \([2]\), and \([3]\) are all scalars which join into the array \([1, 2, 3]\). By contrast, the data \([1, 2]\), \texttt{["3"]} are \emph{not} uniform due to differing dimensions (1x2 vs. 1x1) and data types (double vs. string). 

\begin{figure}[ht]
    \centering
    \includegraphics[width=9cm]{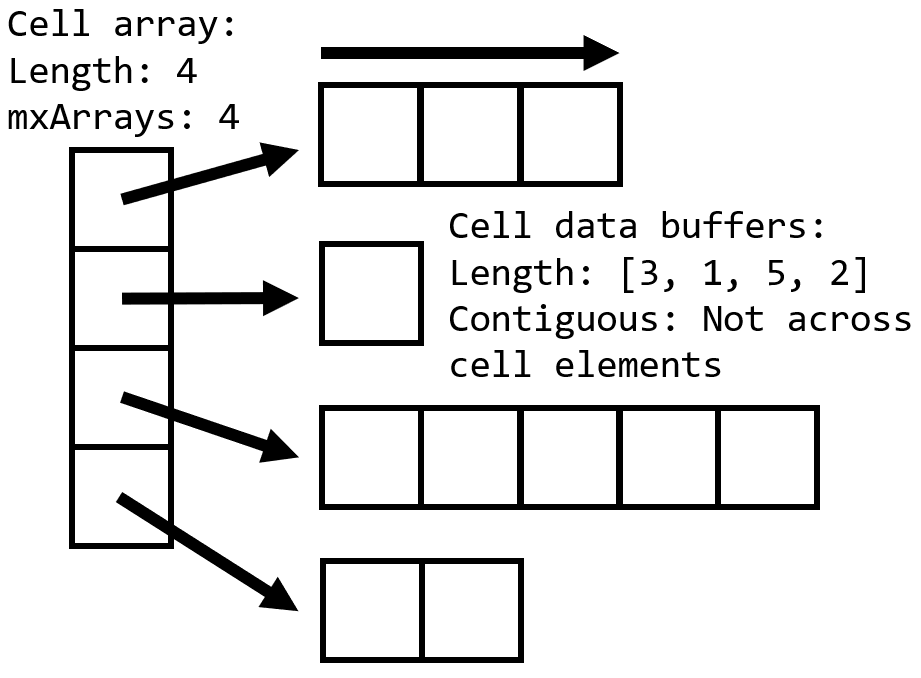}
    \caption{Cell array memory layout in MATLAB}
    \label{fig:MemoryLayout}
\end{figure}

\emph{cellstr} is a non-uniform data container which allows it to hold textual data of differing sizes. Each element is stored in an independent \emph{cell}. These \emph{mxArrays} and their data are chained together in memory, like a linked list or double-ended queue. Although the \emph{mxArrays} themselves are contiguous in memory, the data buffer that each point to is not, as demonstrated in Figure \ref{fig:MemoryLayout}.

Each cell array data pointer points to a disparate memory address. Iterating across a cell array leads to poor data locality and frequent cache-misses, thus degrading performance. Cell arrays are inherently inefficient.

\textbf{Recommendation}: \emph{Use string array over cellstr} to benefit from a simpler syntax, faster execution, and a more compact memory layout.

\section{String concatenation}
String concatenation is the operation of fusing smaller strings into a longer string. Most programming languages support this operation. MATLAB has three distinct ways to concatenate text:

\begin{enumerate}
    \item The \href{https://www.mathworks.com/help/matlab/ref/sprintf.html}{\emph{printf}} family of functions which use format specifiers (\emph{ex: sprintf, fprintf, sscanf}, ...)
    \item \href{https://www.mathworks.com/help/matlab/ref/horzcat.html}{horzcat} which uses brackets, ex: [\textquotesingle hello, \textquotesingle, name]
    \item String \href{https://www.mathworks.com/help/matlab/ref/plus.html}{concatenation} which uses the `+' operator
\end{enumerate}

String concatenating and formatting are fundamental to managing textual data. This foundational requirement is compared analytically below.

\subsection{\emph{printf/sprintf}}

\begin{figure}[ht]
    \centering
    \includegraphics[width=15cm]{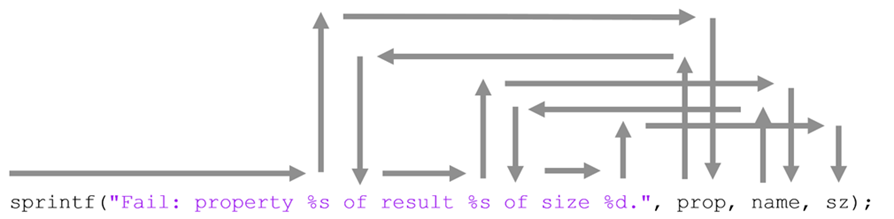}
    \caption{The flow of a \emph{sprintf} formatted string.}
    \label{fig:SprintfFlow}
\end{figure}

\emph{printf} uses format specifiers which act as placeholders in the unformatted text. The corresponding variable substitutions are input arguments to the \emph{sprintf} function. For example,
\begin{verbatim}
    >> sprintf('hello, %s!', name);
\end{verbatim}
The textual variable \texttt{name} is substituted for the \emph{character} format specifier \texttt{\%s} to complete the string. Specifying only one input argument to \emph{sprintf} is simple, but how does substitution scale for multiple inputs? Figure \ref{fig:SprintfFlow} demonstrates a trace of one's eyeballs while reading a string which makes three variable substitutions.

To follow this string from left to right, one's eyes repeatedly need to dart to the end of the expression to locate the variable, make a mental map of its format specifier and variable type, then return to the string. Navigating formatted text requires eyeball gymnastics for all but the simplest text, but fortunately there exists a simpler way: string concatenation.

\subsection{String concatenation}
\begin{figure}[ht]
    \centering
    \includegraphics[width=14cm]{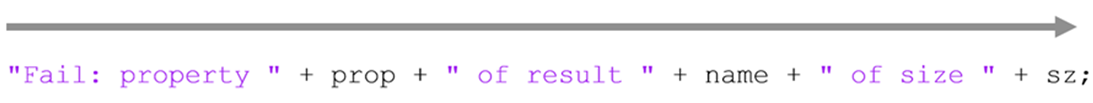}
    \caption{The arrow of a concatenated string.}
    \label{fig:ConcatFlow}
\end{figure}

Figure \ref{fig:ConcatFlow} is a visual trace of the same string above but this time using the `+' (concatenate) operator.

This flows from left to right. In addition, the format specifiers \texttt{\%s} and \texttt{\%d} are not needed: MATLAB already knows variable data types via its tagged union \emph{mxArray} (\S \ref{mxArray}) which makes format specifiers superfluous. This allows MATLAB's string library to convert efficiently non-string values (such as numeric) into string automatically (an \emph{implicit} conversion). These advantages make string concatenation easier to read and write as compared to \emph{sprintf}, but how do their execution times compare?

\subsection{Concatenation performance comparison}
\label{PerformanceComparison}

Here are three MATLAB code samples, each concatenating identical text in a different way:
\begin{verbatim}
    >> sprintf("%d %s", 1, a)
    >> [num2str(1) ' ' a]
    >> 1 + " " + a
\end{verbatim}

All three create the same text content (assume that the variable \(a\) equals the character `a'). These examples test the \emph{double} and \emph{char} data types given that they are among MATLAB's most widely used \cite{MATLABData}. Also, the text is deliberately kept short to focus more heavily on comparing their string concatenating properties. Table \ref{table:Concat} summarizes the performance comparison for each of the three ways to concatenate text.

\begin{table}[ht]
\begin{tabular}{|l|l|l|l|l|}
\texttt{Metric}          & \texttt{sprintf}    & \texttt{{[}num2str(1) ` ' a{]}} & \texttt{1 + " " + a} & \texttt{Advantage over sprintf} \\ \hline
\texttt{Chars of M-code} & \texttt{22 chars}   & \texttt{18 chars}               & \texttt{11 chars}    & \texttt{2x as compact}          \\
\texttt{Duration (sec)}  & \texttt{0.00001375} & \texttt{0.00001227}             & \texttt{0.000001693} & \texttt{8.1x faster}               
\end{tabular}
\caption{Comparison of MATLAB text concatenation.}
\label{table:Concat}
\end{table}

Table \ref{table:Concat} clearly indicates that string concatenation has strong advantages over \emph{printf} for the concatenated text above:
\begin{itemize}
    \item Zero function calls (ex: no \emph{sprintf}) for string concatenation
    \item Faster execution: string operators JIT-compile (see \S \ref{JIT}) better than function calls \cite{Performance}
    \item Fewer characters of MATLAB code, which often increases code clarity
    \item Implicit numeric-to-string conversion (no \emph{num2str} calls needed)
    \item No format specifiers (\texttt{\%d} and \texttt{\%s} are common enough to memorize but rare specifiers such as \texttt{\%x}, \texttt{\%g}, and \texttt{\%E} likely require documentation \cite{sprintf})
    \item No surprise output if a user: (1) accidentally chooses the wrong format specifier (e.g., floating-point instead of integer), or (2) changes the variable data type but forgets to update its specifier
    \item Flow left to right as demonstrated by the arrows in Figures \ref{fig:SprintfFlow} and \ref{fig:ConcatFlow}
\end{itemize}

\subsection{MATLAB's Just-In-Time (JIT) compilation}
\label{JIT}
MATLAB code runs through a JIT compiler \cite{ExecutionEngine}. JIT compilation is a hybrid between compilation ahead-of-time (AOT) and interpretation of a language. In terms of speed, generally \texttt{AOT > JIT > Interpreted}. JIT compilation often falls in the performance spectrum between AOT and Interpreted: efficient JIT compilation nears AOT performance, while inefficient JIT is no better than Interpreted.

One of the principles of JIT is that time-to-compile is generally faster than time-to-execute. This is especially important for tight loops and vectorized instructions. By compiling once \emph{then reusing} that optimized bytecode for subsequent executions, performance greatly benefits \cite{Loren}. MATLAB launched its new execution engine in R2015a which continues to speed up every release: MATLAB benchmarks in R2020a are on average 2.18x faster than pre-R2015a \cite{MATLABPerformance}. To circle back to string vs. cell, MATLAB strings have a greater data to metadata ratio and are inherently vectorized. This allows JIT compilation for \emph{string} more opportunities to optimize. \\

\textbf{Recommendation}: With improved clarity, conciseness, data type conversions, and performance, \emph{prefer string concatenation to printf.}

\section{Autocomplete}
Many MATLAB functions offer autocomplete to improve documentation lookup, reduce syntactic errors, and aid in development time. This ease of use is especially of importance to new users beginning to learn the language. MATLAB is frequently used by engineers and researchers who have strong problem-solving backgrounds but are relatively new to software development.

Autocomplete is yet another aspect where string shines over \emph{cellstr}. To illuminate the difference in autocomplete for these two types, here is another coding exercise: \\

\emph{Retrieve the image number from a list of images names where the number token is separated by an underscore, ex: "1001\_img.jpg" should return "1001".} \\

\begin{figure}[ht]
    \centering
    \includegraphics[width=5cm]{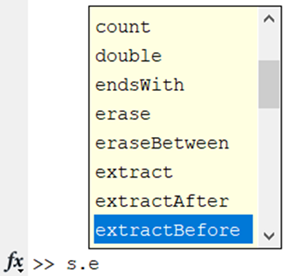}
    \caption{String method autocomplete}
    \label{fig:StringAutocomplete}
\end{figure}

Starting with string array:

\begin{verbatim}
    >> s = ["10_img.jpg", "11_img.jpg"];
\end{verbatim}

To see the autocomplete, type \texttt{"s."} then \texttt{<TAB>}. Figure \ref{fig:StringAutocomplete} shows the hinted string functions.

\begin{figure}[ht]
    \centering
    \includegraphics[width=7cm]{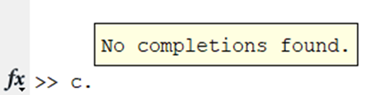}
    \caption{Cell array's empty autocomplete}
    \label{fig:CellAutocomplete}
\end{figure}

After a quick scan, a user finds the function that he or she needs, \emph{extractBefore}:

\begin{verbatim}
    >> s = s.extractBefore("_")
    
s =
  1×2 string array

    "10"    "11"
\end{verbatim}

How does autocomplete for cell arrays compare? Here are the same file names, but this time as a cell array of character vectors:
\begin{verbatim}
    >> c = {'10_img.jpg', '11_img.jpg'};
\end{verbatim}

As before, type the variable name then \texttt{<TAB>}. Figure \ref{fig:CellAutocomplete} shows the lack of helpful results.

Why are there no suggestions? This is because MATLAB's autocomplete for variables works with \emph{methods} rather than \emph{functions}. Cell arrays are a container, not a true class, and therefore have no methods. This makes finding the correct function a larger challenge.

\subsection{Can MATLAB's \href{https://www.mathworks.com/products/matlab/live-editor.html}{Live Editor} help?}
The Live Editor is MATLAB's new and revamped text editor, containing additional features such as block selection, enhanced autocomplete, and formatted text with LaTeX. With MATLAB's Live Editor, autocomplete appears as soon as a user \emph{begins} typing. Let's say that an engineer is looking for string methods beginning with \emph{"ex"}:

\begin{figure}[ht]
    \centering
    \includegraphics[width=6cm]{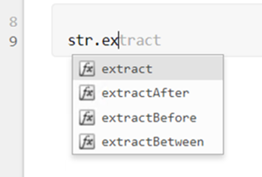}
    \caption{Live Editor's autocomplete}
\end{figure}

The Live Editor's popup menu lists four results, all of which are relevant to text processing. New developers who are unfamiliar with MATLAB's first-party functions immediately find what they are looking for.

\begin{figure}[ht]
    \centering
    \includegraphics[width=9cm]{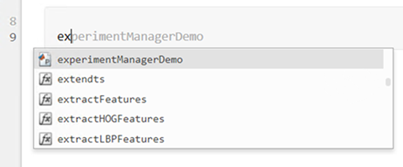}
    \caption{Live Editor's unfiltered autocomplete}
    \label{fig:LiveUnfiltered}
\end{figure}

For cell array, however, users must \emph{begin} by typing \emph{"ex"} rather than scoping into a variable. The Live Editor does not have an opportunity to filter and therefore returns all 200+ candidates as shown in Figure \ref{fig:LiveUnfiltered}.

\emph{extractBefore} is buried in there somewhere but scanning through hundreds of records is not helpful for anyone, regardless of experience level.

\textbf{Recommendation}: \emph{prefer string arrays over cellstr} to let autocomplete improve your development efficiency.

\section{What if MATLAB string arrays are too slow?}
\begin{figure}[ht]
    \centering
    \includegraphics[width=15cm]{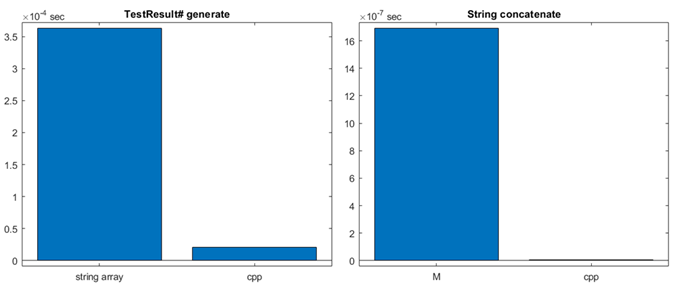}
    \caption{C++ vs. MATLAB duration comparison (lower is better)}
    \label{fig:CppPerf}
\end{figure}

Although MATLAB string arrays are vastly superior to \emph{cellstr} in performance, they still incur the cost of living in a higher-level language. When additional speedup is needed, consider developing in a compiled ahead-of-time language such as C or C++. Figure \ref{fig:CppPerf} illustrates the same two MATLAB performance benchmarks compared with C++: (1) "TestResult\#" string array generation and (2) concatenation. Note: \emph{cellstr} and \emph{sprintf} in MATLAB are so slow that they cause the C++ bar to disappear and are therefore omitted.

\subsection{Google Benchmark}
Google Benchmark is a C++ library for benchmarking. It is particularly suited for micro-benchmarking a critical function which is then pounded by the framework with many thousands of iterations. The library executes a block of C++ code repeatedly until the duration across iterations becomes statistically stable \cite{GoogleBenchmark}.

\subsection{TestResult\# vector generation using Google Benchmark}
Strings are aggressively optimized during compilation and extremely efficient in C++. This makes Google Benchmark the perfect tool for measuring C++ string performance. The C++ \texttt{TestResultArray()} function below creates an array of the same ["TestResult1", "TestResult2", ..., "TestResult1000"] strings as benchmarked earlier. It begins by preallocating a \texttt{vector<string>}. Then the function converts each numeric index to string followed by concatenating it to the current element in the string vector.

\begin{verbatim}
    static void TestResultArray(benchmark::State &state)
    {
        for (auto _ : state) // the 'for' loop defines the scope to measure
        {
            constexpr int numel = 1000;
            vector<string> v(numel, "TestResult");
            for (size_t x = 0; x < v.size(); ++x)
            {
                v[x] += std::to_string(x + 1);
            }
        }
    }
\end{verbatim}

The above C++ code sample uses a loop, but the algorithmic approach shown below is an interesting coding exercise: it uses \texttt{std::iota} to create an incrementing sequence then \texttt{std::transform} with a \emph{lambda} expression to convert each integer index to string:

\begin{verbatim}
    constexpr int numel = 1000;
    vector<string> v(numel, "TestResult");
    array<int, numel> num;
    iota(num.begin(), num.end(), 1);
    transform(num.begin(), num.end(), v.begin(), v.begin(),
        [] (int n, string &s) { return s += std::to_string(n); });
\end{verbatim}

\begin{figure}[ht]
    \centering
    \includegraphics[width=12cm]{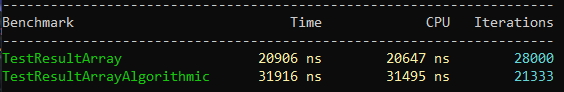}
    \caption{Google Benchmark output for the TestResult benchmarks}
    \label{fig:StringConc}
\end{figure}

The algorithm is about 50\% slower than the loop due to visiting each element multiple times. The loop completes in 20,647 nanoseconds on average while the algorithm takes 31,495 nanoseconds as shown in Figure \ref{fig:StringConc}.

\subsection{String concatenate using Google Benchmark}

The C++ \texttt{StringConcatenate()} function concatenates the same string studied in Section \ref{PerformanceComparison} and is incredibly efficient, taking only 25.1 nanoseconds on average after 24.8 million iterations. It converts the numeric index \(1\) to string then uses the concatenate operator to build the final string. Additionally, it leverages \texttt{DoNotOptimize} to prevent the compiler from optimizing the local string variable out of existence. This ensures that the concatenation action is accurately being measured \cite{DoNotOptimize}.

\begin{verbatim}
    static void StringConcatenate(benchmark::State &state)
    {
        for (auto _ : state)
        {
            char a = 'a';
            std::string s = std::to_string(1) + " " + a;
            benchmark::DoNotOptimize(s);
        }
    }
\end{verbatim}

\begin{figure}[ht]
    \centering
    \includegraphics[width=12cm]{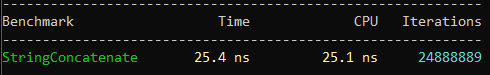}
    \caption{Google Benchmark output for the string concatenate benchmark}
\end{figure}

\section{Hardware and Software}
All results were obtained using MATLAB R2020a on Windows 10, Intel® Xeon® W-2133 CPU @ 3.60GHz, 64 GB RAM.

\section{Conclusion}
Strings are pervasive in data science. Processing volumes of text efficiently is a critical aspect of data analysis. MATLAB provides two textual containers to aid researchers: string array and cell array. Comparing them objectively repeatedly demonstrates the strengths of string over cell. In terms of performance, string arrays are faster than \emph{cellstr} for almost all string processing benchmarks such as string-building, formatting, concatenation, and implicit data type conversions. By memory footprint, string utilizes approximately half as much metadata to hold identical arrays of text content as compared to \emph{cellstr}. Additionally, string arrays offer a cleaner, more natural syntax which improves developer usability by reducing syntax errors and code churn. And lastly, autocomplete powerfully supports string arrays which reduces development time. Given this wealth of advantages, MATLAB users should prefer string arrays over cell array of character vectors and proactively replace existing \emph{cellstr} usages.

\bibliographystyle{abbrv}
\bibliography{references}  

\end{document}